\documentclass[aps,prd,twocolumn,superscriptaddress,nofootinbib]{revtex4-1}

\usepackage{latexsym}
\usepackage{amsmath}
\usepackage{amssymb}
\usepackage{amsfonts}
\usepackage{bm}
\usepackage{physics}

\usepackage{color}
\definecolor{purple}{rgb}{0.5,0,0.5}
\definecolor{blue}{rgb}{0.0,0,0.9}
\definecolor{prdblue}{rgb}{0.133,0.118,0.498}
\usepackage[colorlinks=true, pdfstartview=FitV, linkcolor=prdblue, citecolor= prdblue, urlcolor=prdblue]{hyperref}

\usepackage{supertabular} 
\usepackage{placeins}
\usepackage{epsfig}
\usepackage{graphicx}

\usepackage{soul} 
\usepackage{color}

\usepackage{lipsum}
\usepackage{amsthm}
\usepackage{bbold}
\usepackage{dcolumn}
\usepackage{longtable}
\usepackage{epstopdf}
\usepackage{xspace}
\usepackage{cancel}
\usepackage{nicefrac}
\usepackage{multirow}
\usepackage{float}

\begin{document}

\title{$\mathbf{\gamma^{(*)} + N(940)\frac{1}{2}^+ \to N(1520)\frac{3}{2}^{-}}$ helicity amplitudes and transition form factors}

\author{L. Albino}
\email[]{luis.albino.fernandez@gmail.com}
\affiliation{Departamento Sistemas F\'isicos Qu\'imicos y Naturales, Universidad Pablo de Olavide, Sevilla, 3800708, Spain.}
\affiliation{Departmento de Ciencias Integradas, Universidad de Huelva, E-21071 Huelva, Spain.}

\author{G. Paredes-Torres}
\email[]{gustavo.paredes@umich.mx}
\affiliation{Instituto de F\'{i}sica y Matem\'aticas, Universidad Michoacana de San Nicol\'as de Hidalgo, Morelia, Michoac\'an
58040, M\'{e}xico.}
\affiliation{The Abdus Salam ICTP, Strada Costiera 11, 34151 Trieste, Italy.}
\affiliation{SISSA, via Bonomea 265, 34136, Trieste, Italy.}

\author{K. Raya}
\email[]{khepani.raya@dci.uhu.es}
\affiliation{Departmento de Ciencias Integradas, Universidad de Huelva, E-21071 Huelva, Spain.}

\author{A. Bashir}
\email[]{adnan.bashir@umich.mx}
\affiliation{Departmento de Ciencias Integradas, Universidad de Huelva, E-21071 Huelva, Spain.}
\affiliation{Instituto de F\'{i}sica y Matem\'aticas, Universidad Michoacana de San Nicol\'as de Hidalgo, Morelia, Michoac\'an
58040, M\'{e}xico.}

\author{J. Segovia}
\email[]{jsegovia@upo.es}
\affiliation{Departamento Sistemas F\'isicos Qu\'imicos y Naturales, Universidad Pablo de Olavide, Sevilla, 3800708, Spain.}

\date{\today}

\begin{abstract}
We recently reported new results on the $\gamma^{(*)} + N(940)\frac{1}{2}^+ \to \Delta(1700)\frac{3}{2}^{-}$ transition form factors using a symmetry-preserving treatment of a vector$\,\otimes\,$vector contact interaction (SCI) within a coupled formalism based on the Dyson-Schwinger, Bethe-Salpeter, and Faddeev equations. In this work, we extend our investigation to the $\gamma^{(*)} + N(940)\frac{1}{2}^+ \to N(1520)\frac{3}{2}^{-}$ transition.
Our computed transition form factors show reasonable agreement with experimental data at large photon virtualities. However, deviations emerge at low $Q^2$, where experimental results exhibit a sharper variation than theoretical predictions. This discrepancy is expected, as these continuum QCD analyses account only for the quark-core of baryons, while low photon virtualities are dominated by meson cloud effects. We anticipate that these analytical predictions, based on the simplified SCI framework, will serve as a valuable benchmark for more refined studies and QCD-based truncations that incorporate quark angular momentum and the contributions of scalar and vector diquarks.
\end{abstract}


\maketitle


\section{Introduction}
\label{sec:Intro}

All ordinary matter is composed of atoms, each comprising a dense nucleus of protons and neutrons, collectively known as nucleons. Nucleons belong to a broader family of femtometer-scale particles called hadrons. Research on hadrons has revealed that they are complex bound states of quarks and gluons, which interact through the strong nuclear force. This interaction is described by quantum chromodynamics (QCD) which is a Poincaré-invariant, non-Abelian gauge field theory.

Despite the elegance and apparent simplicity of the QCD Lagrangian, the hadron spectrum does not emerge in an obvious manner~\cite{Brambilla:2014jmp,Gross:2022hyw}. Instead, it arises from highly complex non-perturbative phenomena such as color confinement~\cite{Alkofer:2006fu,Greensite:2003bk}, dynamical chiral symmetry breaking (DCSB) for quarks~\cite{Glozman:2007ek,Mitter:2014wpa}, and the generation of an effective gluon mass scale in the infrared~\cite{Cornwall:1981zr,Ayala:2012pb,Papavassiliou:2022wrb}. These complexities  result in rich hadron spectroscopy, a collection of color singlet hadrons made up of  quarks and gluons. A successful classification scheme for hadrons in terms of up, down and strange quarks and antiquarks was independently proposed by Murray Gell-Mann~\cite{Gell-Mann:1964ewy} and George Zweig~\cite{Zweig:1964CERN} in 1964. It categorizes hadrons as mesons and baryons which are quark-antiquark and three-quark bound-states, respectively, and form multiplets of the $SU(3)$ flavor symmetry.

The excited states of the nucleon, collectively named as $N^\ast$ resonances, provide valuable information on how QCD builds the baryon spectrum. High-luminosity experimental facilities such as the Thomas Jefferson National Accelerator Facility (JLab) in USA~\cite{Dudek:2012vr, Accardi:2023chb}, MAMI and ELSA in Germany~\cite{Tiator:2011pw, Tiator:2018pjq} or J-PARC in Japan~\cite{Aoki:2021cqa} have been designed in order to measure electromagnetic excitation of nucleon resonances, $\gamma^{(*)} + N \rightarrow N^\ast$ to uncover the baryon spectrum and to extract their transition electro-couplings, $g_vNN^\ast$, from the meson electro-production data~\cite{Crede:2013kia}. Obtained primarily with the CLAS detector at the JLab, the electro-couplings of all low-lying $N^\ast$ states with mass less than $1.6\,\text{GeV}$ have been determined via independent analyses of $\pi^+ n$, $\pi^0 p$ and $\pi^+ \pi^- p$ exclusive channels. Moreover, preliminary results for the $g_v N N^\ast$ electro-couplings of most higher-lying $N^\ast$ states with masses below $1.8\,\text{GeV}$ have also been obtained from CLAS meson electro-production data~\cite{CLAS:2009ces, Aznauryan:2012ba, Proceedings:2020fyd}. It is worth highlighting that considerable theoretical effort has been made in parallel with the experimental measurements. An updated review can be found in Ref.~\cite{Ramalho:2023hqd}, where the interested reader can explore the related context and original works.

A vast array of observables related to hadrons can be computed by solving the relevant Dyson-Schwinger equations (DSEs), whose solutions are subsequently used to formulate covariant bound-state equations, including the Bethe-Salpeter equation for mesons and the Faddeev equation for baryons which describe dressed quark-antiquark and three-quark systems, respectively. This framework naturally unifies the infrared and ultraviolet behavior of hadronic observables as its mathematical construction makes no recourse to the strength of the strong interaction. Furthermore, it provides a direct connection between the fundamental degrees of freedom in QCD, quarks and gluons, and the experimentally measurable properties of color-singlet hadrons~\cite{Roberts:1994dr, Maris:2003vk, Eichmann:2016yit, Qin:2020rad}. Nevertheless, obtaining physically reliable solutions requires significant systematic effort and continuous refinement~\cite{Bashir:2012fs}. Over the years, remarkable progress in this approach has enabled it to contribute successfully even to precision observables within the standard model of particle physics~\cite{Raya:2019dnh, Miramontes:2021exi}.

A key result which stems from the studies of DSEs with realistic quark-quark interactions~\cite{Qin:2011dd, Binosi:2014aea} is the natural emergence of nonpointlike quark-quark (diquark) correlations within baryons~\cite{Maris:2002yu, Eichmann:2008ef, Cloet:2011qu, Eichmann:2016yit, Barabanov:2020jvn, Paredes-Torres:2024mnz}. Growing empirical evidence supports the presence of diquark correlations in the proton~\cite{Close:1988br, Cloet:2005pp, Cates:2011pz, Segovia:2015ufa, Cloet:2012cy, Cloet:2014rja}. It is important to emphasize that these correlations differ markedly from the elementary diquarks introduced about fifty years ago to simplify the three-quark bound-state problem~\cite{Lichtenberg:1967zz, Lichtenberg:1968zz}. Modern studies using pseudo-Faddeev equations predict dynamic, two-body correlations in which dressed quarks participate in all diquark clusters. Notably, the baryon spectrum obtained through the quark-diquark picture shows substantial agreement with predictions from three dressed-quark frameworks, lattice QCD calculations, and experimental results (see Refs.~\cite{Eichmann:2016yit, Barabanov:2020jvn} for reviews).

The quark-diquark model of baryons, combined with symmetry-preserving quark-quark interaction kernels and vertices that either follow QCD-like momentum dependence or employ simplified contact interactions, has proven particularly effective in describing the electro-couplings of low-lying nucleon resonances, including the corresponding ground states: $N(940)\frac{1}{2}^{+}$~\cite{Wilson:2011aa, Cloet:2008re, Segovia:2014aza, Cui:2020rmu, Yao:2024uej} and  $\Delta(1232)\frac{3}{2}^{+}$,  see~\cite{Segovia:2013rca, Segovia:2013uga, Segovia:2014aza, Segovia:2016zyc}; their first radial excitations: $N(1440)\frac{1}{2}^{+}$~\cite{Wilson:2011aa, Segovia:2015hra, Chen:2018nsg} and $\Delta(1600)\frac{3}{2}^{+}$~\cite{Lu:2019bjs}; and their parity partners: $N(1535)\frac{1}{2}^{-}$~\cite{Raya:2021pyr} and $\Delta(1700)\frac{3}{2}^{-}$~\cite{Albino:2025fcp}. 

As part of our ongoing effort to compute the transition form factors of nucleon resonances, we study the process $\gamma^{(*)} + N(940)\frac{1}{2}^+ \to N(1520)\frac{3}{2}^-$. This transition is particularly interesting because a symmetry-preserving contact interaction (SCI) treatment of the DSEs describes both the $N(1520)\frac{3}{2}^-$ and $\Delta(1700)\frac{3}{2}^-$ baryons through the same single Faddeev amplitude, associated with an isovector--axial-vector diquark correlation. Notably, recent studies~\cite{Liu:2022ndb, Liu:2022nku} using a continuum approach to the three-valence-quark bound-state problem in relativistic quantum field theory—where the Faddeev equation kernel and interaction vertices exhibit QCD-like momentum dependence—demonstrate that this Faddeev amplitude dominates the internal dynamics of $\Delta(1700)\frac{3}{2}^-$ overwhelmingly, contributing $99.98\%$ of its mass. The same amplitude accounts for $91.60\%$ of the mass of the $N(1520)\frac{3}{2}^-$. Having recently provided a theoretical description of the $\gamma^{(*)} + N(940)\frac{1}{2}^+ \to \Delta(1700)\frac{3}{2}^-$ transition, it seems both natural and straightforward to extend this analysis to the $\gamma^{(*)} + N(940)\frac{1}{2}^+ \to N(1520)\frac{3}{2}^-$ process. Furthermore, our results can be compared with experimental measurements at various photon virtualities~\cite{Burkert:2002zz, CLAS:2009tyz, CLAS:2009ces, CLAS:2012wxw, Mokeev:2015lda,  Mokeev:2023zhq, ParticleDataGroup:2024cfk}, providing insight into the internal structure of the $N(1520)\frac{3}{2}^{-}$ resonance. 

Although the symmetry-preserving contact interaction (SCI) within the DSEs approach captures only the infrared dynamics of relativistic bound-state systems, it still retains crucial QCD features such as Poincar\'e invariance, quark confinement by ensuring the absence of quark production
thresholds, adequate amount of dynamical chiral symmetry breaking, conservation of axial vector and vector Ward identities as well as the Golberger-Treiman relations. Our findings geberally highlight a noticeable sensitivity of $N^\ast$ electro-couplings to the internal structure of the involved baryons, offering valuable insight. Despite its inherent limitations, the SCI framework’s algebraic simplicity allows us to make analytical predictions that serve as a useful benchmark, providing a valuable guide for more sophisticated studies within QCD-akin DSE framework, which we plan to adopt in near future.

The manuscript is organized as follows. Sec.~\ref{sec:Theory} is devoted to the fundamental aspects of our theoretical framework, emphasizing the similarities and differences of the Faddeev amplitude for both the $N(1520)\frac{3}{2}^-$ and $\Delta(1700)\frac{3}{2}^-$ baryons. Sec.~\ref{sec:EFFs} provides a brief description of the electromagnetic interactions which contribute to the elastic as well as the transition form factors. It also presents the general decomposition of the electromagnetic current for the $N(940)\frac{1}{2}^+$, $N(1520)\frac{3}{2}^-$ and the $N(940)\frac{1}{2}^+$-to-$N(1520)\frac{3}{2}^-$ process expressed in terms of electromagnetic form factors. We also provide the relation between helicity amplitudes and transition form factors. In Sec.~\ref{sec:Results}, we discuss our numerical results for both transition form factors and helicity amplitudes, comparing them with experimental measurements. Finally, we provide a brief summary of our findings and an outlook for future in Sec.~\ref{sec:Summary}.


\begin{figure}[!t]
\centerline{%
\includegraphics[clip,width=0.40\textwidth, height=0.09\textheight]{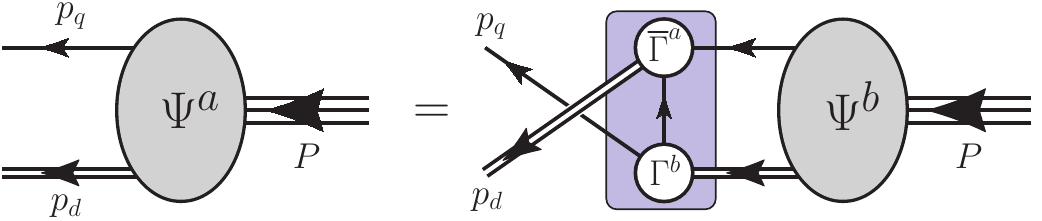}}
\caption{\label{fig:Faddeev} Poincar\'e covariant Faddeev equation.  $\Psi$ is the Faddeev amplitude for a baryon of total momentum $P= p_q + p_d$.  The shaded rectangle demarcates the kernel of the Faddeev equation: \emph{single line}, dressed-quark propagator; $\Gamma$,  diquark correlation amplitude; and \emph{double line}, diquark propagator.}
\end{figure}

\section{Baryon Faddeev amplitude}
\label{sec:Theory}

Our description of baryon bound-states is based on a Poincar\'e-covariant Faddeev equation, illustrated in Fig.~\ref{fig:Faddeev}. Its key elements are the dressed-quark and diquark propagators, and the diquark Bethe-Salpeter amplitudes. All quantities are fully determined once the quark-quark interaction kernel is specified. As explained in the introduction, we use a symmetry-preserving regularization of a vector$\,\otimes\,$vector contact interaction~\cite{Yin:2019bxe, Yin:2021uom}. This interaction is characterized by a constant gluon propagator which is consistent with the infrared (IR) finite behavior of the gluon propagator observed in lattice regularized QCD~\cite{Bowman:2004jm, Bogolubsky:2009dc, Aguilar:2008xm}. Moreover, the effective coupling derived from this approach aligns well with contemporary estimates of the infrared behavior of QCD's running-coupling~\cite{Binosi:2016nme, Cui:2019dwv}. We then incorporate this interaction within a rainbow-ladder (RL) truncation of the DSEs, utilizing the tree-level quark-gluon vertex to ensure that the quark propagator exhibits a momentum-independent dressed-quark mass. In this study, we adopt the set of model parameters from Ref.~\cite{Yin:2021uom}, as our goal is to provide consistent predictions constrained by previously established results.

Once the dressed-quark propagator, dressed-diquark propagators and canonically normalized diquark Bethe-Salpeter amplitudes are etermined~\cite{Yin:2019bxe, Yin:2021uom}, the baryon's wave function in the quark-diquark picture can be compactly expressed as
\begin{eqnarray}
\Psi = \Psi^1 +\Psi^2 +\Psi^3 \,,
\end{eqnarray}
where the superscript stands for the spectator quark, \emph{e.g.} $\Psi^{1,2}$ are obtained from $\Psi^3$ by a cyclic permutation of all the quark labels. As demonstrated in Ref.~\cite{Liu:2022ndb}, within a framework based on a Faddeev equation kernel and interaction vertices exhibiting QCD-like momentum dependence, the $\Delta(1700)\frac{3}{2}^-$ baryon is predominantly governed by isovector-axial-vector diquark correlations. In Ref.~\cite{Albino:2025fcp}, we assumed that a simple but realistic representation of the Faddeev amplitude of a positively-charged $\Delta(1700)\frac{3}{2}^-$ is given by
\begin{eqnarray}
\label{eq:DeltaFA}
\Psi_\Delta^3 = \sum_{j=1,2} \Gamma^{1^{+}_j}_{\alpha}(p_1,p_2) \Delta^{1^{+}}_{\alpha\beta}(K) \mathcal{D}^{j}_{\beta\rho}(P) u_{\rho}(P) \,, \label{Delta(1700) Faddeev Amplitude}
\end{eqnarray}
where the flavor structure has been omitted for the sake of simplicity in the notation (see Ref.~\cite{Chen:2012qr} for a more comprehensive overview on this issue), and the matrix $\mathcal{D}^{j}_{\beta\rho}(P)$ describes the quark-diquark momentum correlation of the $\Delta$-baryon and, within the SCI-approach, simplifies to
\begin{equation}
\label{eq:quarkdiquarkDelta}
\mathcal{D}^{j}_{\beta\rho}(P) = d^{1^{+}_j} \, \delta_{\beta\rho} \,,
\end{equation}
with $d^{\left\{ ud \right\}}=\sqrt{2} d^{\left\{ uu \right\}} = \sqrt{2/3}$, provided the two possible isovector-axial-vector diquarks with flavour content $1^{+}_1 = \left\{ uu \right\}$ and $1^{+}_2 = \left\{ ud \right\}$. The remaining terms shown in Eq.~\eqref{eq:DeltaFA} are detailed in~\cite{Albino:2025fcp}: $\Gamma^{1^{+}_j}_{\alpha}(p_1,p_2)$, $\Delta^{1^{+}}_{\alpha\beta}(K)$ and $u_{\rho}(P)$ are, respectively, the Bethe-Salpeter isovector--axial-vector diquark amplitude, isovector--axial-vector diquark propagator and Rarita-Schwinger spinor representing an on-shell $\Delta$-baryon.

Note that the QCD-kindred analysis in Ref.\,\cite{Liu:2022nku} predicts a subdominant but measurable contribution (around $20\%$) from the scalar $0^+$ correlation to the wave function normalization of the $N(1520)\frac{3}{2}^-$ baryon. However, since the SCI approach forbids the presence of relative momenta in the baryon’s Faddeev amplitude, a closer examination of  Eqs.(4) and~(5) in Ref.~\cite{Liu:2022nku} reveals that the only SCI-allowed quark-diquark momentum correlations in the $N(1520)\frac{3}{2}^-$ Faddeev amplitude are those corresponding to isoscalar-vector and isovector-axial-vector. If the first one is neglected, as the analysis in Ref~\cite{Liu:2022nku} shows that the isoscalar-vector diquark contributes a small fraction ($\sim 3.4 \%$) to the $N(1520)\frac{3}{2}^-$ Faddeev amplitude, it follows that the Faddeev amplitudes of the $N(1520)\frac{3}{2}^-$ and $\Delta(1700)\frac{3}{2}^-$ are identical. 

The argument above indicates that, within the SCI-approach, the Faddeev amplitude of the $N(1520)\frac{3}{2}^-$ can be directly obtained from that of the $\Delta(1700)\frac{3}{2}^-$, differing only by an overall normalization constant which produces
\begin{equation}
m_{N(1520)\frac{3}{2}^-} = 0.89 \times m_{\Delta(1700)\frac{3}{2}^-} \,, \label{N(1520) mass}
\end{equation}
and a slight variation in the flavor structure,
\begin{equation}
d^{\left\{ uu \right\}}= -\sqrt{2} d^{\left\{ ud \right\}} \,, 
\end{equation}
where $d^{\left\{ ud \right\}} = -1/\sqrt{3}$. It arises purely from Clebsch-Gordan coefficients that describe the coupling between isovector--axial-vector diquarks and either an isospin-$1/2$ $N(1520)\frac{3}{2}^-$ baryon or an isospin-$3/2$ $\Delta(1700)\frac{3}{2}^-$ one.

There are two technical aspects that warrant explanation here. The first aspect concerns  Faddeev amplitude of the nucleon, $N(940)\frac{1}{2}^+$, since this baryon is involved in the description of the electromagnetic transition $\gamma^{(\ast)} + N(940)\frac{1}{2}^+\to N(1520)\frac{3}{2}^-$. We use the same amplitude as reported in~\cite{Albino:2025fcp}, \emph{i.e.} the Faddeev amplitude of a positively charged nucleon given by
\begin{eqnarray}
\label{Nucleon FA}
\Psi^3 &=& \Gamma^{0^{+}}(p_1,p_2)\Delta^{0^{+}}(K)\mathcal{S}(P)u(P) \nonumber \\
&& \hspace{-.5cm} +\sum_{j=1,2} \Gamma^{1^{+}_j}_{\alpha}(p_1,p_2)\Delta^{1^{+}}_{\alpha\beta}(K)\mathcal{A}^j_{\beta}(P)u(P) \,,
\end{eqnarray}
where the quantities $\Gamma^{0^{+}}(p_1,p_2)$, $\Delta^{0^{+}}(K)$ and $u(P)$ are all defined in~\cite{Albino:2025fcp} and correspond, respectively, to the Bethe-Salpeter scalar diquark amplitude, scalar diquark propagator and a spinor satisfying the Dirac equation $\left(i\gamma \cdot P + m_B\right)u(P) = 0$ for an on-shell nucleon with momentum $P$.

The second technical aspect is related with the fact that the Faddeev kernel (represented by the shaded rectangle in Fig.~\ref{fig:Faddeev}) involves diquark breakup and reformation via exchange of a dressed-quark. We follow Ref.~\cite{Roberts:2011cf} and make a marked but convenient simplification with impunity, see Ref.~\cite{Xu:2015kta}; namely, in the Faddeev equation for a baryon of type $B$, the quark exchanged between the diquarks is represented as
\begin{equation}
S^T=\frac{g_B^2}{M} \,,
\end{equation}
where the superscript $T$ indicates matrix transpose, $M$ is the dressed quark mass and $g_B$ is an effective coupling. This is a variant of the so-called \emph{static approximation}, which was introduced in Ref.~\cite{Buck:1992wz} and has subsequently been used in studies of a wide range of baryon properties~\cite{Yin:2019bxe, Yin:2021uom}.


\begin{figure}[!t]
\centering
\includegraphics[scale=0.60]{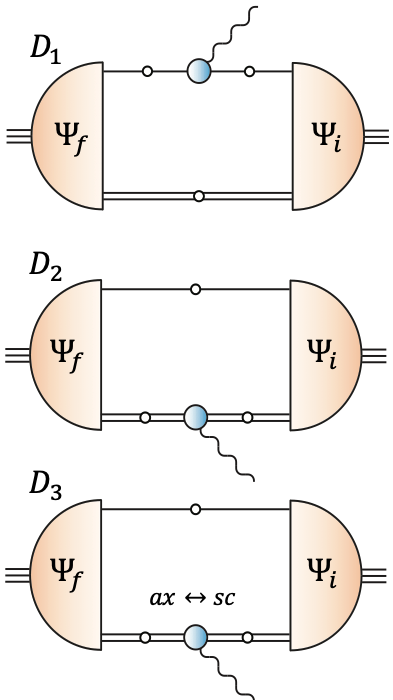}
\caption{\label{fig:EMInteractions} Diagrammatic representation of contributions for elastic and transition EM currents in the quark-diquark picture. Dressed-quark and diquark propagators are represented by single and double lines, respectively. Initial ($\Psi_i$) and final ($\Psi_f$) state's Faddeev amplitude for the involved baryons are represented by orange semi-circles. The blue blobs represent the corresponding quark-photon and diquark-photon vertices for each of the following diagrams: in $D_1$ it entails the photon coupling to a dressed-quark; in $D_2$ the photon couples elastically to a diquark; and $D_3$ involves photon-induced transitions between axial-vector and scalar diquarks.}
\end{figure}

\section{The photon-baryon interaction}
\label{sec:EFFs}

In our theoretical framework, calculating the elastic form factors for both the $N(940)\frac{1}{2}^+$ and $N(1520)\frac{3}{2}^-$ baryons is essential. This is due to the fact that the value of the leading electric form factor at $Q^2 = 0$ is required to properly normalize the Faddeev amplitudes. Such normalizations are subsequently used in the analysis of the $\gamma^{(*)} + N(940)\frac{1}{2}^+ \to N(1520)\frac{3}{2}^-$ transition, which is empirically accessible~\cite{Burkert:2002zz, CLAS:2009tyz, CLAS:2009ces, CLAS:2012wxw, Mokeev:2015lda, Mokeev:2023zhq, ParticleDataGroup:2024cfk}.

The elastic and transition form factors of interest can be derived from the following microscopic current: 
\begin{eqnarray}
\mathcal{J}^{B_f B_i}_{\mu \left[ \rho \right] \left[ \sigma \right]}(P_f,P_i) && \nonumber \\
&& \hspace{-2cm} = \hspace{-.2cm} \sum_{n=1,2,3} \int_{dk}{ \mathcal{P}^{B_f}_{ \left[ \rho \alpha \right]} \mathcal{G}_f^{\pm} \mathcal{G}_{f/i} \Lambda^{D_n}_{\mu \left[ \alpha \right] \left[ \beta \right]} \left(k; P_f, P_i\right) \mathcal{G}_i^{\pm} \mathcal{P}^{B_i}_{ \left[ \beta \sigma \right]} } \,, \nonumber \\
\end{eqnarray}
where $B_{i(f)}$ denotes the initial and final baryon states, with incoming and outgoing momenta $P_i$ and $P_f$, respectively. Moreover, the integration notation entails $\int_{dk} = \int d^4k/(2\pi)^4$ and the sum is performed over the possible quark-photon and diquark-photon contributions $\Lambda^{D_n}_{\mu \left[ \alpha \right] \left[ \beta \right]} \left(k; P_f, P_i\right)$. Within the SCI approach and only considering isoscalar-scalar and isovector-axial-vector diquark correlations, there are three types of contributions that ensure current conservation. These are diagrammatically represented in Fig.~\ref{fig:EMInteractions}; the corresponding mathematical content has been recently described in Ref.~\cite{Albino:2025fcp}.

Furthermore, baryon's parity is reflected via the Dirac structures $\mathcal{G}^{+(-)}_{i,f} = \mathbf{I}_{\mathrm{D}} (\gamma_5)$, the additional $\mathcal{G}_{f/i}$ is equal to $\gamma_5$ for a flipping parity transition between initial and final baryons and it is $\mathbf{I}_{\mathrm{D}}$ for baryon's parity conservation in the transition. Besides, Lorentz indices denote photon and $N(1520)\frac{3}{2}^-$ baryon polarizations: $\mu$ is associated with the photon whereas Greek letters between square brackets indicate that the indices will only appear if they are associated with the appearance of $N(1520)\frac{3}{2}^-$ in either the initial or final state. For nucleon, the operator $\mathcal{P}^B_{\left[ \rho \alpha \right]} (P)$ reduces to the positive-energy projector $\Lambda_{+}(P) = (1-i \gamma \cdot \hat{P})/2$, with $\hat{P} = P/m_N$ normalized to the nucleon mass. For the $N(1520)\frac{3}{2}^-$, the operator $\mathcal{P}^B_{\left[ \rho \alpha \right]} (P)$ is again the positive energy projector, changing $m_N=1.14\,\text{GeV}$ by $m_{N^\ast} = m_{N(1520)\frac{3}{2}^-}$, and extended by the Rarita-Schwinger projection operator $\mathcal{R}_{\rho \alpha} (P)$, i.e.,
\begin{eqnarray}
\mathcal{P}^N (P) & = & \Lambda_{+}(P) \,, \\
\mathcal{P}^{N^\ast}_{\rho \alpha} (P) & = & \Lambda_{+}(P) \mathcal{R}_{\rho \alpha} (P) \,,
\end{eqnarray}
where
\begin{eqnarray}
\mathcal{R}_{\rho \alpha} (P) & = & \delta_{\rho \alpha} -\frac{1}{3} \gamma_{\rho} \gamma_{\alpha} \nonumber \\
&& +\frac{2}{3} \hat{P}_{\rho} \hat{P}_{\alpha} +\frac{1}{3} \left( \gamma_{\rho} \hat{P}_{\alpha} - \gamma_{\alpha} \hat{P}_{\rho} \right) \,.
\end{eqnarray}
It is worth emphasizing that the microscopic current for the $\gamma^{(*)} + N(940)\frac{1}{2}^+ \to N(1520)\frac{3}{2}^-$ transition can be summarized as
\begin{eqnarray}
\mathcal{J}_{\mu, \rho} \left( P_f, P_i \right) \hspace{-1mm} &=& \hspace{-1mm} \left( d^{\left\{ uu \right\}} e_{d} + d^{\left\{ ud \right\}} e_{u} \right) \mathcal{J}_{\mu, \rho}^{D_1} \left( P_f, P_i \right) \nonumber \\
&+& \hspace{-1mm} \left( d^{\left\{ uu \right\}} e_{\left\{ uu \right\}} + d^{\left\{ ud \right\}} e_{\left\{ ud \right\}} \right) \mathcal{J}_{\mu, \rho}^{D_2} \left( P_f, P_i \right) \nonumber \\
&+& \hspace{-1mm} d^{\left\{ ud \right\}} e_{\left\{ ud \right\}} \mathcal{J}_{\mu, \rho}^{D_3} \left( P_f, P_i \right) \,, \label{microscopic current}
\end{eqnarray}
in terms of the $D_i$-type ($i=1,2,3$) contributions represented by each diagram, $D_i$, sketched in Fig.~\ref{fig:EMInteractions}. The factorization shown above isolates the similarities and differences between the nucleon-to-$N(1520)\frac{3}{2}^-$ and nucleon-to-$\Delta(1700)\frac{3}{2}^-$ transitions; namely, both are described by the same microscopic contributions, ${\cal J}_{\mu,\rho}^{D_i}$, but the distinct values of the corresponding Faddeev coefficients, $d^{1^{+}_j}$, weight them differently, thus leading to distinct transition profiles.

On the other hand, the macroscopic current that describes the photon-induced transition from $N(940)\frac{1}{2}^+$ to $N(1520)\frac{3}{2}^-$ may be represented as 
\begin{eqnarray}
\mathcal{J}_{\mu, \rho} \left( P_f, P_i \right) =  i \mathcal{P}^{N^\ast}_{\rho \alpha}  \hspace{-.1cm} \left( P_f \right) \Gamma_{\mu, \alpha} \left( P_f, P_i \right) \Lambda_{+} \hspace{-.1cm} \left( P_i \right) \,, \label{Transition current}
\end{eqnarray}
where the decomposition for $\Gamma_{\mu, \alpha} \left( P_f, P_i \right)$ can be written in terms of three Jones-Scadron form factors: the magnetic dipole, $G^{*}_M$, the electric quadrupole, $G^{*}_E$, and the Coulomb quadrupole, $G^{*}_C$:
\begin{eqnarray}
\Gamma_{\mu, \alpha} \left( P_f, P_i \right) &=& \nonumber \\
&& \hspace{-2cm} b \left[ -\frac{i \omega}{2 \lambda_{+}} \left( G^{*}_M - G^{*}_E \right) \mathcal{V}^1_{\alpha \mu} - G^{*}_E \mathcal{V}^2_{\alpha \mu} + \frac{i \tau}{\omega} G^{*}_C \mathcal{V}^3_{\alpha \mu} \right] \,, \nonumber \\ \label{Transition vertex}
\end{eqnarray}
where
\begin{eqnarray}
\lambda_{+} &=& \frac{1+\sqrt{1-4\delta^2}}{2} + \tau \,, \\
\omega &=& \sqrt{\delta^2 + \tau \left( 1 + \tau \right)} \,, \\
b &=& \sqrt{\frac{3}{2}} \left( 1+ \frac{m_{N^\ast}}{m_N} \right) \,.
\end{eqnarray}
with $\tau = Q^2 \hspace{-.05cm}/ \hspace{-.1cm} \left[ 4 \bar{m}^2 \right]$, $\delta = \left( m_{N^\ast}^2-m_N^2\right)\hspace{-.1cm} / \hspace{-.1cm}\left[ 4 \bar{m}^2 \right]$ and  $\bar{m}^2 = \left( m_N^2 + m_{N^\ast}^2 \right)\hspace{-.1cm}/2$. Moreover, the tensors $\mathcal{V}^i_{\alpha \mu} \equiv \mathcal{V}^i_{\alpha \mu} \left( K, Q \right)$ are defined as ($2 K \equiv P_f + P_i$)
\begin{eqnarray}
\mathcal{V}^1_{\alpha \mu} \left( K, Q \right) &=&  \gamma_5 \epsilon_{\alpha \mu \sigma \rho} \mathcal{K}_{\sigma} \left( K,Q \right) \hat{Q}_{\rho} \,, \\
\mathcal{V}^2_{\alpha \mu} \left( K, Q \right) &=& \left( \delta_{\alpha \sigma} - \hat{Q}_{\alpha} \hat{Q}_{\sigma}\right) \bar{\mathcal{T}}_{\sigma \mu} \left( K, Q\right) \,, \\
\mathcal{V}^3_{\alpha \mu} \left( K, Q \right) &=&  \hat{Q}_{\alpha} \mathcal{K}_{\mu} \left( K,Q \right) \,,
\end{eqnarray}
with $\hat{Q} = Q/\left[ 2 \bar{m} \tau \right]$ and
\begin{eqnarray}
\mathcal{K}_{\sigma} \left( K, Q \right) &\equiv& \frac{\sqrt{\tau}}{i \bar{m} \omega} \left( K_{\sigma} + \frac{\delta}{2 \tau} Q_{\sigma} \right) \,, \\
\bar{\mathcal{T}}_{\sigma \mu} \left( K, Q \right) &\equiv& \delta_{\sigma \mu} - \mathcal{K}_{\sigma} \left( K, Q \right)  \mathcal{K}_{\mu} \left( K, Q \right) \,.
\end{eqnarray}
In order to compare directly with experimental measurements, it is often necessary to compute the so-called helicity amplitudes. For the $\gamma^{(*)} + N(940)\frac{1}{2}^+ \rightarrow N(1520)\frac{3}{2}^{-}$ transition, they are expressed in terms of the Jones-Scadron form factors as follows~\cite{Ramalho:2023hqd}
\begin{eqnarray}
\mathcal{A}_{1/2} \left( Q^2 \right) &=& -\frac{1}{4 \mathcal{F}} \left[ G^{*}_{E} \left( Q^2 \right) - 3 G^{*}_{M} \left( Q^2 \right) \right] \,, \label{Helicity Amplitude A12} \\
\mathcal{A}_{3/2} \left( Q^2 \right) &=& -\frac{\sqrt{3}}{4 \mathcal{F}} \left[ G^{*}_{E} \left( Q^2 \right) + G^{*}_{M} \left( Q^2 \right) \right] \,, \label{Helicity Amplitude A32} \\
\mathcal{S}_{1/2} \left( Q^2 \right) &=& -\frac{1}{\sqrt{2} \mathcal{F}} \frac{|\mathbf{q}|}{2 m_{N^\ast}} G^{*}_{C} \left( Q^2 \right)\,, \label{Helicity Amplitude S12} 
\end{eqnarray}
with $|\bm{q}|=\sqrt{Q_+^2Q_-^2}/(2m_{N^\ast})$, $Q_{\pm}^2=(m_{N^\ast}\pm m_N)^2+Q^2$, and
\begin{eqnarray}
\mathcal{F} = \frac{1}{\sqrt{2 \pi \alpha}} \frac{m_N}{m_{N^\ast} - m_N} \sqrt{\frac{m_N \left( m^2_{N^\ast} - m^2_N \right)}{\left( m_{N^\ast} + m_N\right)^2 +Q^2 }} \,,
\end{eqnarray}
with $\alpha = 1/137$. 


\section{RESULTS}
\label{sec:Results}

A detailed analysis of our numerical results for the transition form factors, followed by a discussion of the corresponding helicity amplitudes, describing the $\gamma^{(*)} + N(940)\frac{1}{2}^+ \to N(1520)\frac{3}{2}^-$ reaction, will be presented below. Before that, a few preliminary remarks are in order.

The first comment concerns the SCI-DSE framework and its parameters. It is important to appreciate that the parameters used in this work were already constrained in previous studies. Thus, no additional parameters are introduced here for the analysis of the transition form factors. The second comment addresses our theoretical uncertainties, which fall into two categories: those intrinsic to the numerical algorithm and those related to the parameter determination process. The numerical error is negligible. As mentioned earlier, the model parameters are tuned to reproduce a small set of hadron observables within an acceptable level of agreement with available experimental data. Therefore, assigning specific uncertainties to these parameters -- or to the observables derived from them -- would not be particularly meaningful. In any case, to reasonably assess the parameter uncertainty in our calculations, the numerical results are displayed as a colored band representing the variation in strength, $\eta$, of the dressed-quark anomalous magnetic moment (AMM) which becomes relevant for the photon probes~\cite{Albino:2025fcp}. Specifically, the solid line corresponds to $\eta = 0$, the dotted line to $\eta = 2/3$, and the dashed line indicates the central value at $\eta = 1/3$. This exercise provides insight into how the transition form factors are influenced by the infrared enhancement of the quark AMM, which occurs through dynamical chiral symmetry breaking.

\begin{figure}[!t]
\centering
\includegraphics[scale=.25]{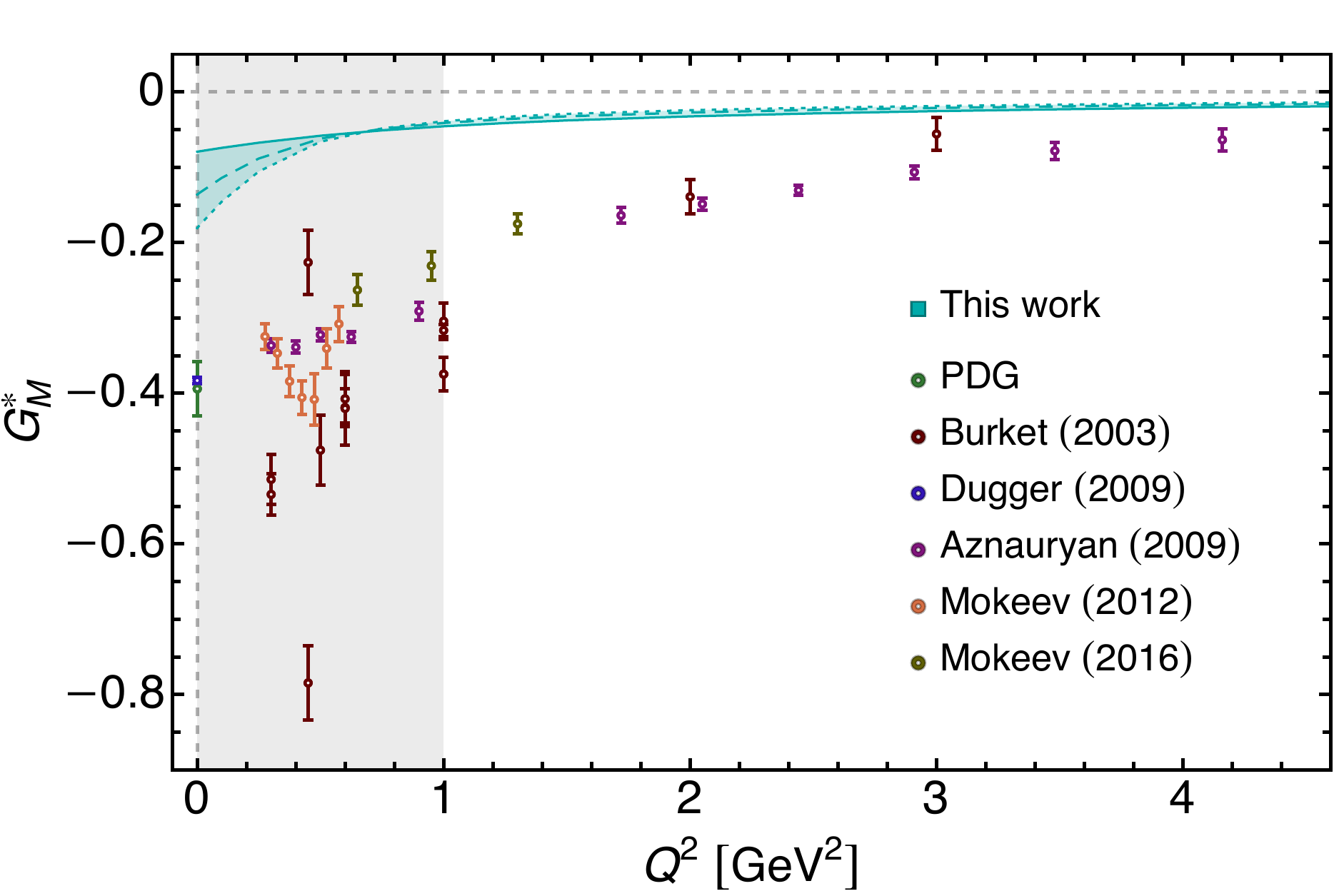}
\caption{\label{fig:GM} Magnetic dipole ($G^{*}_M$) transition form factor as a function of $Q^2$. The data points correspond to experimental measurements taken from Refs.~\cite{Burkert:2002zz, CLAS:2009tyz, CLAS:2009ces, CLAS:2012wxw, Mokeev:2015lda, Mokeev:2023zhq, ParticleDataGroup:2024cfk}. The cyan band represents our numerical results obtained for different values of the parameter $\eta$ that modulates the AMM: $\eta = 0$ (solid line), $\eta = 1/3$ (dashed line), and $\eta = 2/3$ (dotted line). The gray shaded area rectangle covers $Q^2=(0-1)$ GeV$^2$ region where the meson cloud effect is expected to contribute most significantly.}
\end{figure}

The magnetic dipole transition form factor, $G_M^{*}$, is shown in Fig.~\ref{fig:GM} as a function of $Q^2$. The cyan band represents our numerical results within the SCI approach, with its width reflecting variations in the model parameter $\eta \in \left[ 0, 2/3 \right]$. At small values of $Q^2$, our results are relatively far removed from the experimental values. At higher momentum transfers, where the meson cloud effect is expected to fizzle out and the quark core becomes dominant, our computed result is gradually more aligned with the experimental measurements. However, despite being able to produce this form factor with a comparable order of magnitude to the experimental data, the overall disagreement underscores the need for further investigation of the $N(1520)\frac{3}{2}^-$ baryon with momentum-dependent dynamical chiral symmetry breaking. 

\begin{figure}[!t]
\centering
\includegraphics[scale=.25]{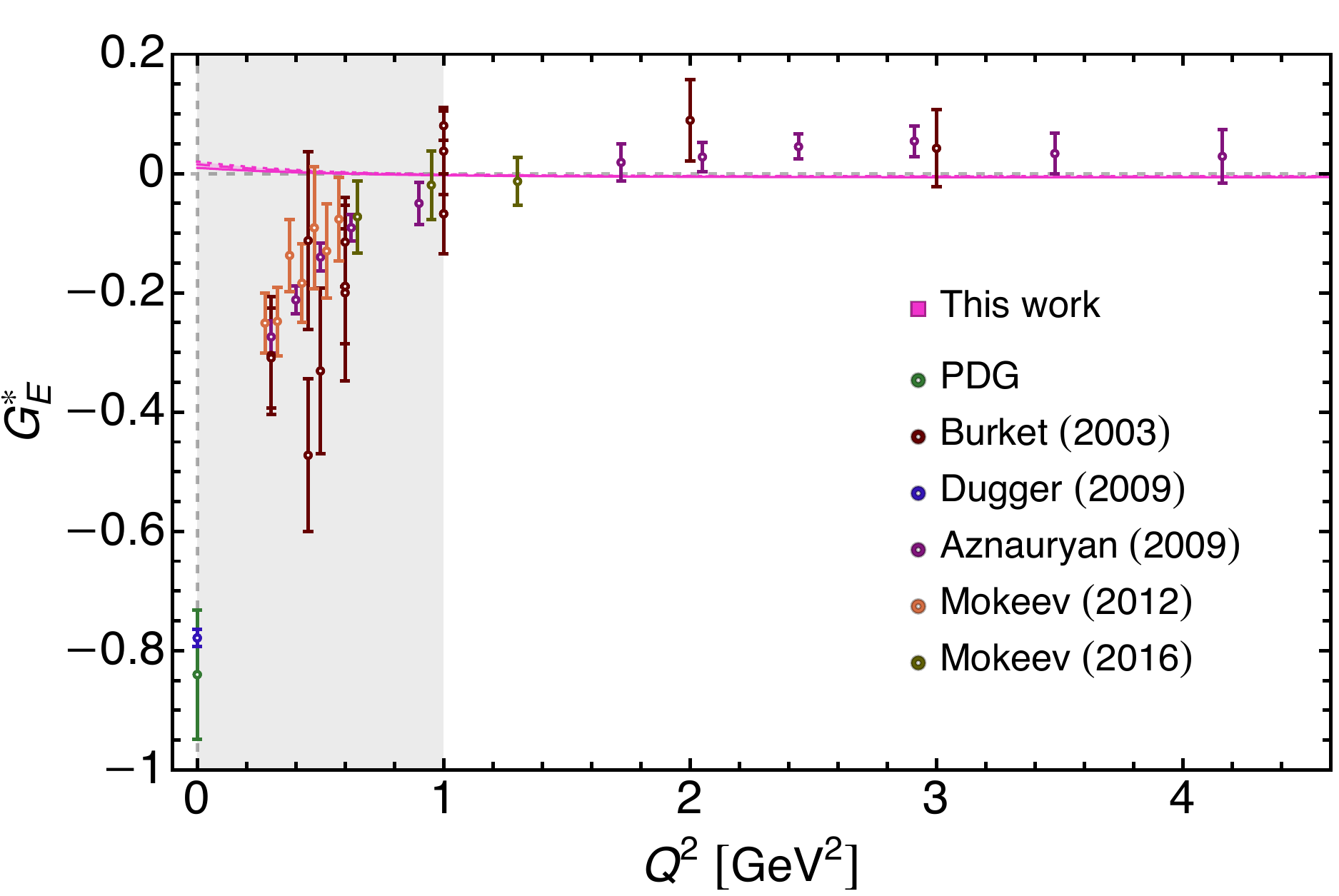}
\caption{\label{fig:GE} Electric quadrupole ($G^{*}_E$) transition form factor as a function of $Q^2$. The data points correspond to experimental measurements taken from Refs.~\cite{Burkert:2002zz, CLAS:2009tyz, CLAS:2009ces, CLAS:2012wxw, Mokeev:2015lda, Mokeev:2023zhq, ParticleDataGroup:2024cfk}. The magenta band represents our numerical results obtained for different values of the parameter $\eta$ that modulates the AMM: $\eta = 0$ (solid line), $\eta = 1/3$ (dashed line), and $\eta = 2/3$ (dotted line). The gray shaded area rectangle covers $Q^2=(0-1)$ GeV$^2$ region where the meson cloud effect is expected to contribute most significantly.}
\end{figure}

Fig.~\ref{fig:GE} shows our results for the electric quadrupole transition form factor $G_E^{*}$ as a function of $Q^2$. The thin magenta band represents the variation in the $\eta$-parameter within the range $\eta \in \left[ 0, 2/3 \right]$. At large $Q^2$, our theoretical curve is compatible with zero and thus it is consistent with the trend observed in the experimental data. Again, the low $Q^2$-region ($Q^2 < 1\,\text{GeV}^2$) is expected to be dominated by the meson cloud effect, superimposed with higher orbital angular momentum components. In fact, any realistic description in this region would require the inclusion of explicit meson degrees of freedom to faithfully account for the meson cloud effect. 

\begin{figure}[!t]
\centering
\includegraphics[scale=.25]{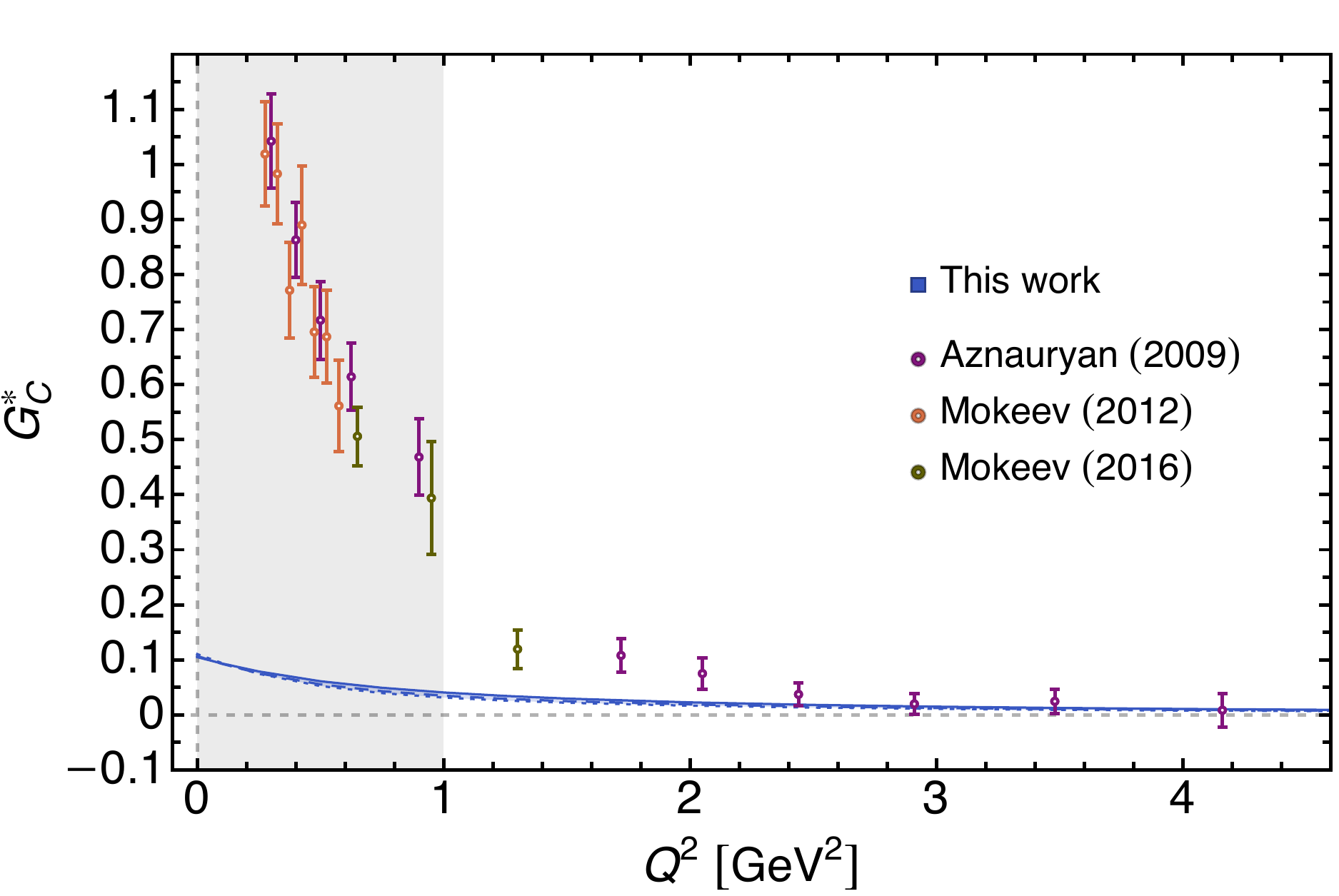}
\caption{\label{fig:GC} Coulomb quadrupole ($G^{*}_C$) transition form factor as a function of $Q^2$. The data points correspond to experimental measurements taken from Refs.~\cite{Burkert:2002zz, CLAS:2009tyz, CLAS:2009ces, CLAS:2012wxw, Mokeev:2015lda, Mokeev:2023zhq, ParticleDataGroup:2024cfk}. The blue band represents our numerical results obtained for different values of the parameter $\eta$ that modulates the AMM: $\eta = 0$ (solid line), $\eta = 1/3$ (dashed line), and $\eta = 2/3$ (dotted line). The gray shaded area rectangle covers $Q^2=(0-1)$ GeV$^2$ region where the meson cloud effect is expected to contribute most significantly.}   
\end{figure}

Finally, our result for the Coulomb quadrupole transition form factor $G_C^{*}$ is presented in Fig.~\ref{fig:GC}, and compared with the available experimental data. The blue band displays our numerical outcome for $\eta \in \left[ 0, 2/3 \right]$. 
Notably, there is no experimental data available for $Q^2 \lesssim 0.5~\text{GeV}^2$.  As anticipated, our computed values remain consistently smaller than the experimental results for $Q^2 < 1\,\text{GeV}^2$  where the meson cloud effect is expected to play a significant role. However, at larger $Q^2$, where this effect diminishes, the improved agreement with experimental data becomes increasingly evident. 

\begin{figure}[!t]
\centering
\includegraphics[scale=.25]{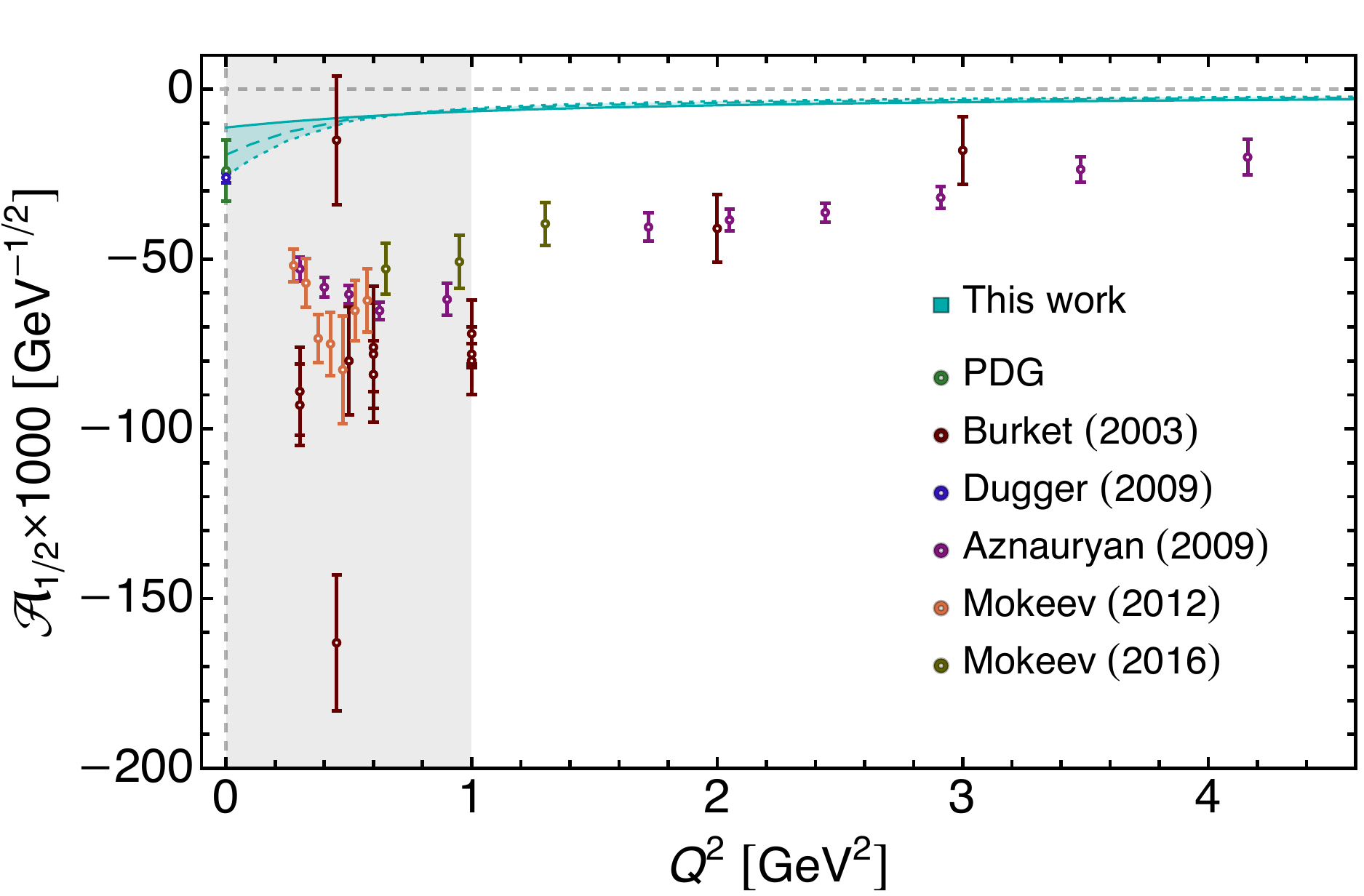}
\includegraphics[scale=.25]{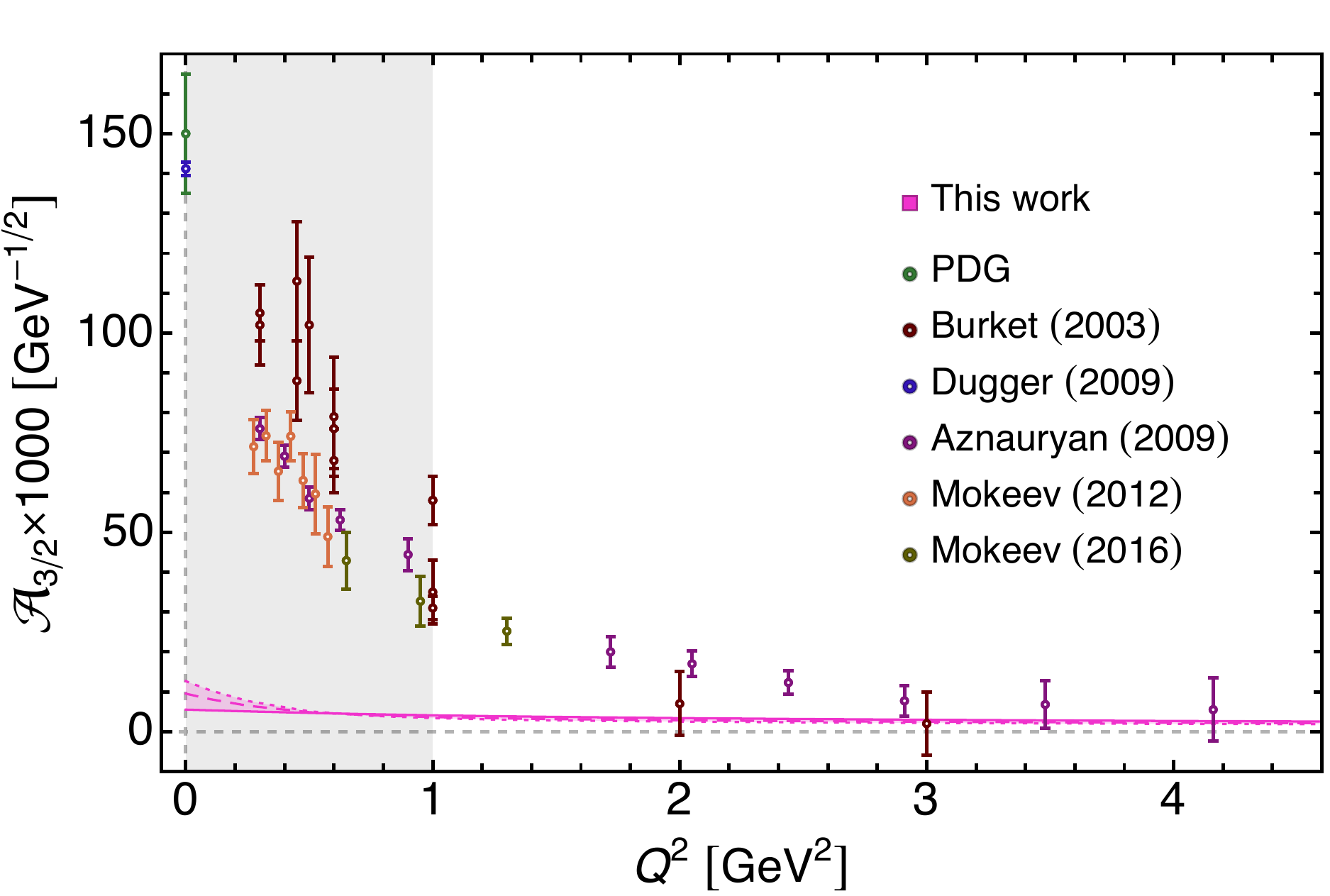}
\includegraphics[scale=.25]{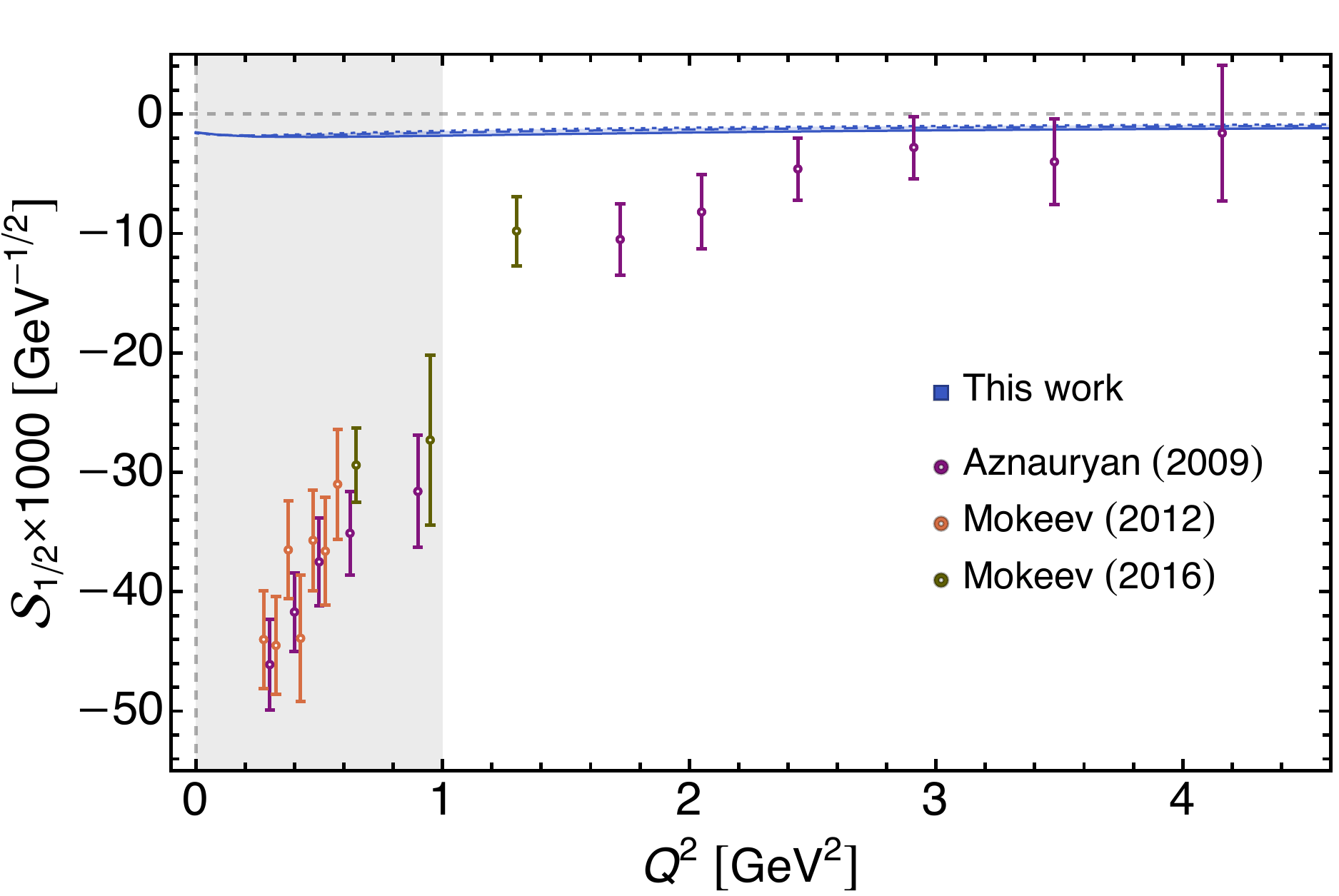}
\caption{\label{fig:helicities} Helicity amplitudes $\mathcal{A}_{1/2}$ (top panel), $\mathcal{A}_{3/2}$ (middle panel) and $\mathcal{S}_{1/2}$ (bottom panel) as a function of $Q^2$. In each panel, the band represents our numerical results obtained for different values of the parameter $\eta$ that modulates the AMM: $\eta = 0$ (solid line), $\eta = 1/3$ (dashed line), and $\eta = 2/3$ (dotted line). The experimental data is taken from Refs.~\cite{Burkert:2002zz, CLAS:2009tyz, CLAS:2009ces, CLAS:2012wxw, Mokeev:2015lda, Mokeev:2023zhq, ParticleDataGroup:2024cfk}. The gray shaded area rectangle covers $Q^2=(0-1)$ GeV$^2$ region where the meson cloud effect is expected to contribute most significantly.}
\end{figure}

Fig.~\ref{fig:helicities} presents analogous results for the helicity amplitudes $\mathcal{A}_{1/2}$ (top panel), $\mathcal{A}_{3/2}$ (middle panel) and $\mathcal{S}_{1/2}$ (bottom panel), as a function of $Q^2$, that describes the $\gamma^{(\ast)} + N(940)\frac{1}{2}^+\to N(1520)\frac{3}{2}^-$ transition. These results are obtained from the transition form factors $G_{M}^{*}$, $G_{E}^{*}$ and $G_C^\ast$ by using Eqs.~\eqref{Helicity Amplitude A12} to~\eqref{Helicity Amplitude S12}. From the top panel of Fig.~\ref{fig:helicities}, one can see that our computed values for $\mathcal{A}_{1/2}$ exhibit a weaker enhancement at lower $Q^2$ compared to the rapid rise observed in the experimental results. This discrepancy can be attributed to the meson cloud effect, momentum-dependent chiral symmetry breaking, a richer diquark composition, and higher orbital angular momentum components. 
The helicity amplitude $\mathcal{A}_{3/2}$, shown in the middle panel of Fig.~\ref{fig:helicities}, seems to be in agreement with experimental data for $Q^2\gtrsim2\,\text{GeV}^2$ but clearly underestimates experimental data at low $Q^2$. In the bottom panel of Fig.~\ref{fig:helicities}, we show our result for $\mathcal{S}_{1/2}$. It matches the sign of the experimental data but remains at lower values compared to the data for entire domain of $Q^2$ spanned. This behavior was anticipated due to the discrepancy observed in the Coulomb transition form factor.

\begin{figure}[!t]
    \centering
    \includegraphics[scale=.25]{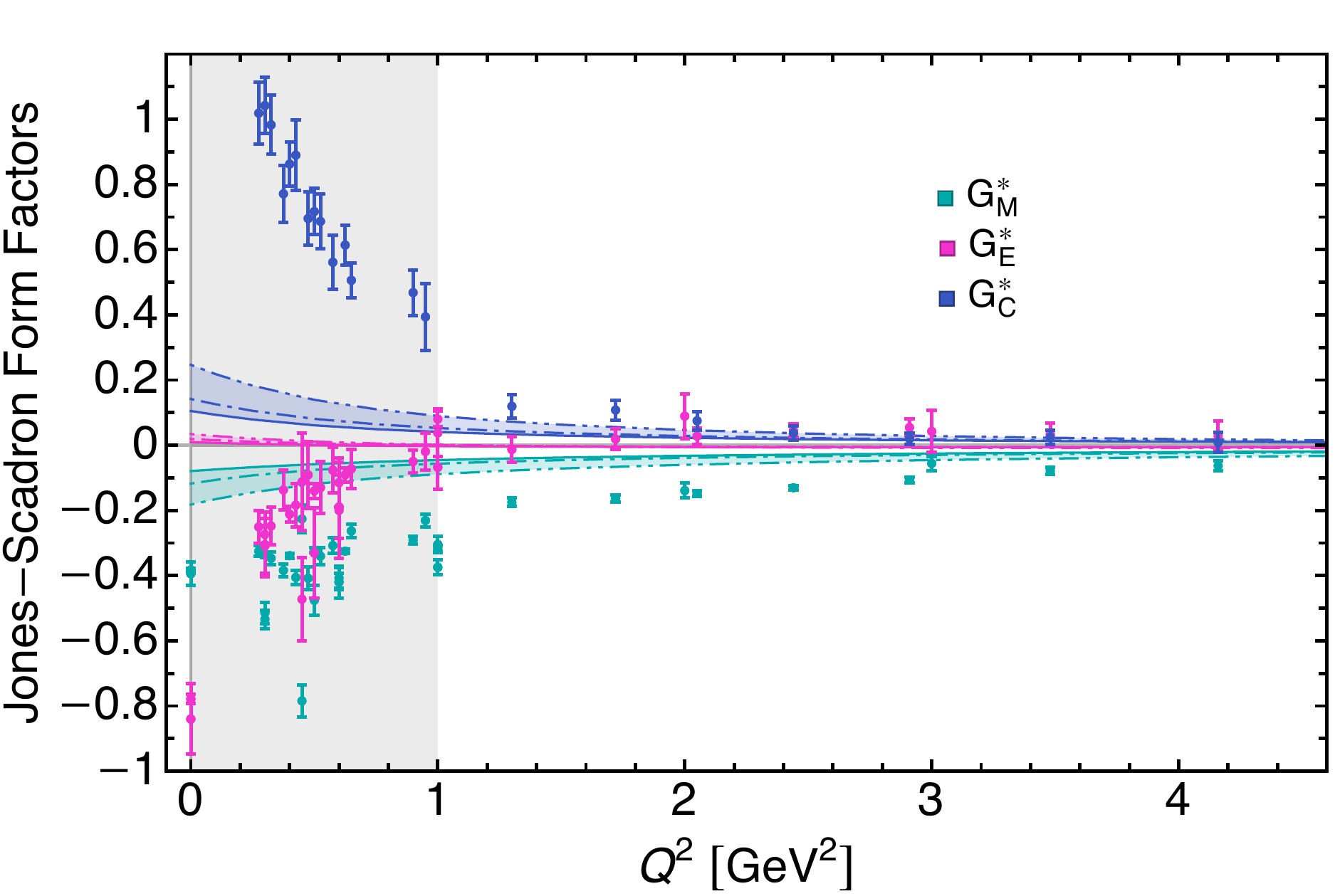}
        \centering
    \includegraphics[scale=.25]{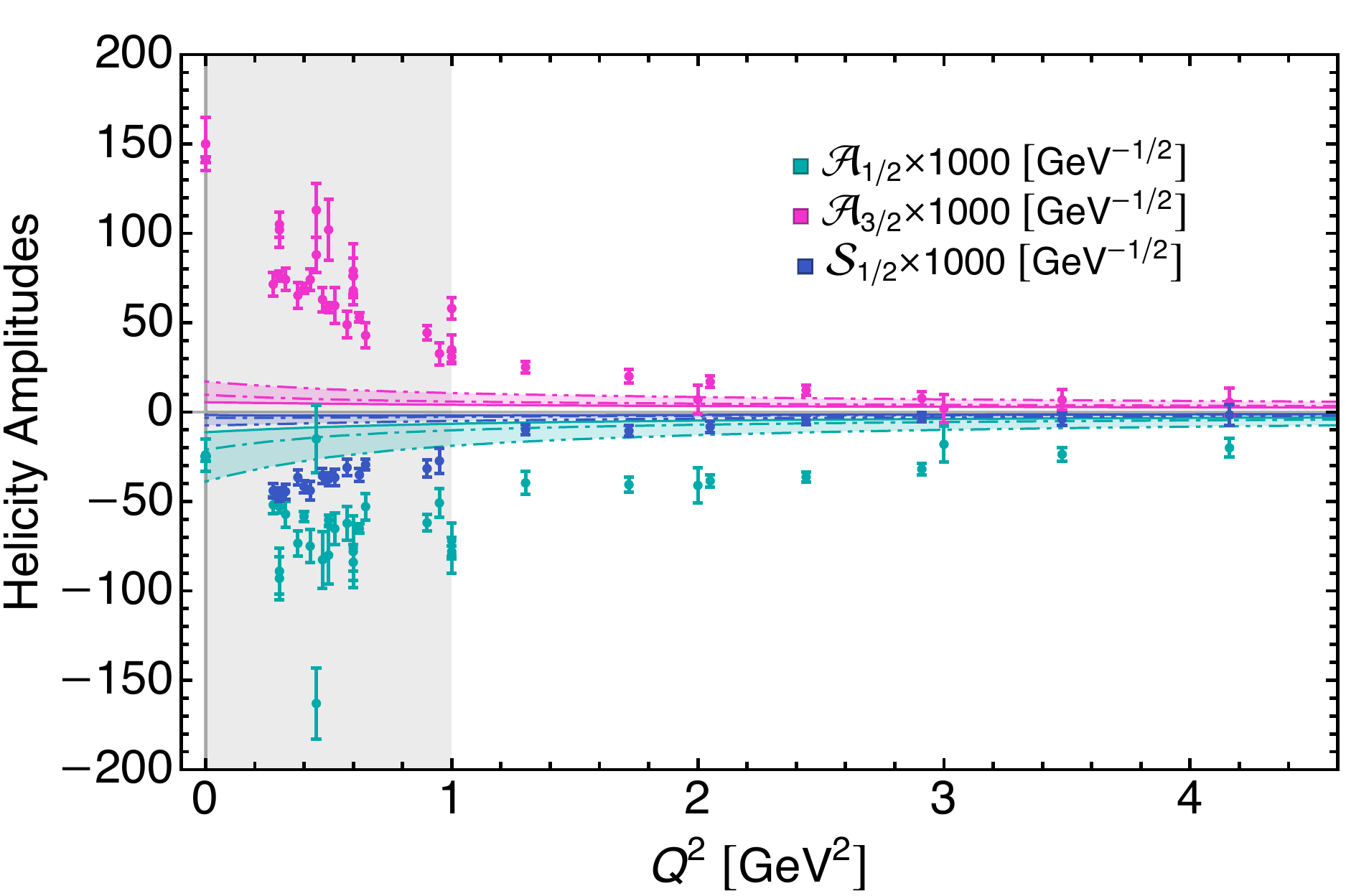}
    \caption{\label{Jones-Scadron and Helicity Amplitudes Variation} Qualitative analysis of the sensitivity of electromagnetic transition form factors under a variation of the $N(1520)$ baryon mass for $\eta = 0$. Top panel - Sensitivity of the Jones-Scadron Form factors $G_M^*$, $G_E^*$ and $G_C^*$. Lower panel - Sensitivity of the helicity amplitudes $\mathcal{A}_{1/2}$, $\mathcal{A}_{3/2}$ and $\mathcal{S}_{1/2}$. For each colored band in both panels, the solid line represents the results for $m_{N(1520)} = 1.54 \, \text{GeV}$, the dashed-dotted line corresponds to $m_{N(1520)} = 1.72 \, \text{GeV}$, and the dashed-double-dotted line uses $m_{N(1520)} = 1.90 \, \text{GeV}$.}
\end{figure}

As a complementary extension of our analysis, we now illustrate the sensitivity of the transition form factors to variations in the mass of the $N(1520)$, in the absence of a dressed-quark anomalous magnetic moment, \emph{i.e.}, for the case $\eta = 0$. This may provide insight into the effects of the meson cloud. Recall that our  results for the Jones-Scadron form factors, displayed in Figs.~\ref{fig:GM}--\ref{fig:GC}, and the helicity amplitudes, shown in Fig.~\ref{fig:helicities}, are computed for $m_{N(1520)} = 1.54 \, \text{GeV}$, \textit{cf.} Eq.~(\ref{N(1520) mass}). Figure~\ref{Jones-Scadron and Helicity Amplitudes Variation} compare our original results with those obtained by enlarging the $N(1520)$ baryon mass $10\%$ and $20\%$, respectively. The solid line represents $m_{N(1520)} = 1.54 \, \text{GeV}$, the dashed-dotted line corresponds to $m_{N(1520)} = 1.72 \, \text{GeV}$, and the dashed-double-dotted line uses $m_{N(1520)} = 1.90 \, \text{GeV}$. Our results show that the form factors become more infrared (IR) enhanced as the $N(1520)$ baryon mass increases. However, they exhibit a convergence to the same values at high $Q^2$. Note also that a similar behavior is expected for different non-zero values of the AMM.

\begin{figure}[!t]
    \centering
    \includegraphics[scale=.25]{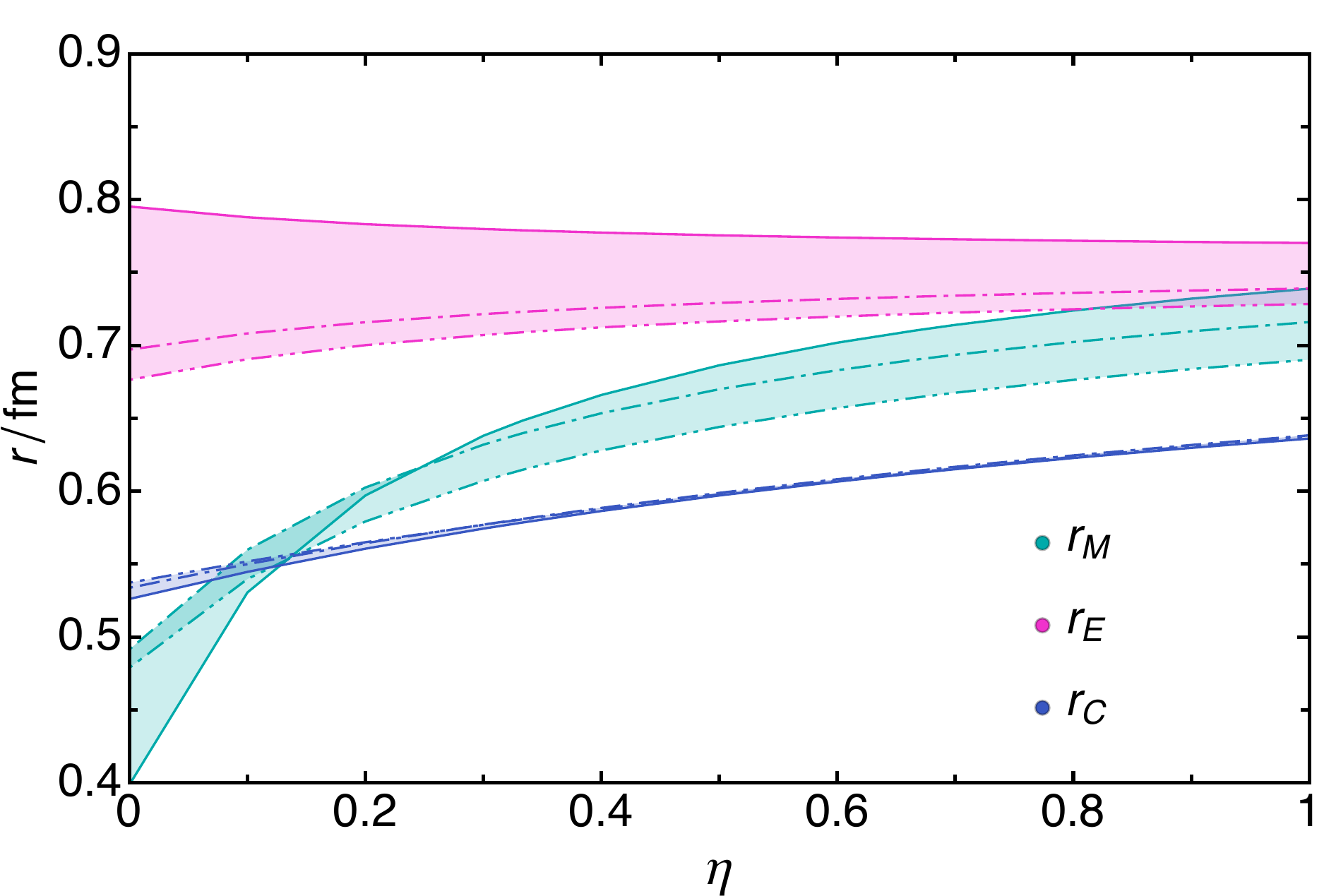}
    \caption{\label{Radii}{The magnetic dipole ($r_M$), electric quadrupole ($r_E$), and Coulomb quadrupole ($r_C$) square-root of mean square radii as a function of the dressed-quark anomalous magnetic moment. For each colored band, the solid line represents the results for $m_{N(1520)} = 1.54 \, \text{GeV}$, the dashed-dotted line corresponds to $m_{N(1520)} = 1.72 \, \text{GeV}$, and the dashed-double-dotted line stands for $m_{N(1520)} = 1.90 \, \text{GeV}$.}}
\end{figure}

We also investigate the effect of the dressed-quark anomalous magnetic moment on the transition form factors for different values of $m_{N(1520)}$ by computing the mass and the AMM sensitivity of the corresponding magnetic ($r_M$), electric ($r_E$), and Coulomb quadrupole ($r_C$) mean square radii,
\begin{eqnarray}
r^2_{\mathcal{F}} =  - \frac{6}{\mathcal{F}(0)} \left. \frac{d \mathcal{F}(Q^2)}{d Q^2} \right|_{Q^2 = 0} \,,
\end{eqnarray}
for the associated form factors $\mathcal{F} = G^{*}_{M}, \, G^{*}_{E}, \, G^{*}_{C}$, respectively. In Fig.~\ref{Radii}, we show the sensitivity of these computed radii to the dressed-quark AMM. Our results suggest a gradual convergence to a narrower band for each radius as $\eta$ increases. Moreover, in Tab.~\ref{Radii Table}, we report the explicit radii values evaluated at the central value $\eta = 1/3$, with uncertainties estimated from the corresponding variation between $\eta = 0$ and $\eta = 2/3$. Notably, our findings exhibit a decreasing trend for the electric radius $r_E$, when the $N(1520)$ baryon mass is higher, physically consonant with a more compact spatial distribution of the electric charge. This increasing behavior is also observed in the magnetic radius $r_M$, except for $m_{N(1520)} = 1.54 \, \text{GeV}$ in $0 \leq \eta \lesssim 0.3$, which may be a consequence of such a low mass adjusted for $N(1520)$ in the SCI framework. In contrast, the Coulomb radius $r_C$ remains essentially constant, suggesting reduced sensitivity of the longitudinal transition structure to changes in the $N(1520)$ mass.

\begin{table}[!t]
\centering
\renewcommand{\arraystretch}{1.5}
\begin{tabular}{|c|c|c|c|}
\hline
 \; $m_{N(1520)}/$GeV \; & 1.54 & 1.72 & 1.90 \\
\hline
$r_{M}$/fm & \; $0.65_{-0.25}^{+0.06}$ \; & \; $0.64_{-0.15}^{+0.05}$ \; & \; $0.61_{-0.14}^{+0.05}$ \; \\
$r_{E}$/fm & $0.78_{-0.01}^{+0.02}$ & $0.72_{-0.03}^{+0.01}$ & $0.71_{-0.03}^{+0.01}$ \\
$r_{C}$/fm & $0.58_{-0.05}^{+0.03}$ & $0.58_{-0.05}^{+0.03}$ & $0.58_{-0.04}^{+0.03}$ \\
\hline
\end{tabular}
\label{Radii Table}
\caption{The magnetic dipole ($r_M$), the electric quadrupole ($r_E$), and the Coulomb quadrupole ($r_C$) square-root of mean square radii as a function of the anomalous magnetic moment of the dressed quark (error bars) and the mass of the $N(1520)$ baryon.}
\end{table}


\section{SUMMARY}
\label{sec:Summary}

This work presents a detailed analysis of the numerical results for the transition form factors and helicity amplitudes associated with the electro-production process $\gamma^{(*)} + N(940)\frac{1}{2}^+ \to N(1520)\frac{3}{2}^-$ in a quark-diquark picture of baryons. We use a symmetry-preserving contact interaction (SCI) within the Dyson-Schwinger Equations (DSEs) framework. 
The model provides a unified prediction that remains fully consistent with the parameters constrained in our earlier study of the $\gamma^{(*)} + N(940)\frac{1}{2}^+ \to\Delta(1700)\frac{3}{2}^-$ transition. By calculating transition form factors and helicity amplitudes, the study aims to shed light on the internal structure of these baryons and serve as a benchmark for future QCD-based studies.
By computing transition form factors and helicity amplitudes, this study aims to illuminate the internal structure of these baryons and serve as a benchmark for future QCD-based investigations.

The main findings focus on the magnetic dipole ($G_M^*$), electric quadrupole ($G_E^*$), and Coulomb quadrupole ($G_C^*$) transition form factors. 
All three align reasonably well with experimental data at large photon virtualities. 
 However, deviations arise at low $Q^2$, where experimental results show a sharper behavior compared to theoretical predictions, consistent with the fact that meson cloud effect dominates this region. The SCI framework is also expected to overlook finer details of the internal dynamics in the involved baryons, such as contributions from quark orbital angular momentum and diquark composition.

  Helicity amplitudes ${\cal A}_{1/2}$, ${\cal A}_{3/2}$, and ${\cal S}_{1/2}$, derived from these form factors, further illustrate the model's performance and validate our conclusions. While the SCI framework captures key qualitative features of these transitions, subtle quantitative discrepancies emphasize the need for more refined truncations.
Despite its limitations, the algebraic simplicity of the SCI model offers valuable benchmark for future studies. These results are expected to support experimental programs at facilities like JLab 12 GeV, and its potential future 22 GeV upgrade, while also motivating investigations of higher resonance states in QCD's non-perturbative regime.


\begin{acknowledgements}
L.A. acknowledges financial support provided by Ayuda B3 ``Ayudas para el desarrollo de l\'\i neas de investigaci\'on propias" del V Plan Propio de Investigaci\'on y Transferencia 2018-2020 de la Universidad Pablo de Olavide, de Sevilla.
G.P.T. acknowledges financial support provided by the National Council for Humanities, Science and Technology (CONAHCyT), Mexico, through their program: {\em Beca de Posgrado en México}.
A.B. acknowledges financial support provided by the {\em Coordinaci\'on de la Investigaci\'on Cient\'ifica} of the {\em Universidad Michoacana de San Nicol\'as de Hidalgo}, Morelia, Mexico, grant no. 4.10, the {\em Consejo Nacional de Humanidades, Ciencias y Tecnolog\'ias}, Mexico, project CBF2023-2024-3544 as well as the Beatriz-Galindo support during his  scientific stay at the University of Huelva, Huelva, Spain. 
Otherwise, this work has been partially financed by 
Ministerio Español de Ciencia, Innovación y Universidades under grant No. PID2022-140440NB-C22;
Junta de Andalucía under contract Nos. PAIDI FQM-370 and PCI+D+i under the title: “Tecnologías avanzadas para la exploración del universo y sus componentes" (Code AST22-0001).
\end{acknowledgements}


\bibliography{preprint_CI_gNN32m}

\end{document}